\documentstyle[12pt]{article}
\begin{document}
\parskip 10pt plus 1pt
\title{Spacetime Dependent Lagrangians and Weak-Strong Duality : Sine Gordon
and Massive Thirring Models}
\date{}
\maketitle
\centerline{\it Rajsekhar Bhattacharyya $^{a}$ and Debashis Gangopadhyay $^{b}$
\footnote{e-mail:debashis@boson.bose.res.in}}
\centerline {$^a$ Department of Physics, Dinabandhu Andrews College, Calcutta-700084, INDIA}
\centerline{$^b$ S.N.Bose National Centre For Basic Sciences,} 
\centerline {JD-Block,Sector-III,Salt Lake, Calcutta-700091, INDIA.}
\baselineskip=20pt

\begin{abstract}

The formalism of spacetime dependent lagrangians developed in Ref.1 is 
applied to the Sine Gordon and massive Thirring models. It is shown
that the well-known equivalence of these models (in the context of
weak-strong duality) can be understood in this approach from the same
considerations as described in Ref.1 for electromagnetic duality.
A further new result  is  that all
these can be naturally linked to the fact that the holographic principle
has analogues at length scales much larger than quantum gravity.
There is also the possibilibity of {\it noncommuting coordinates} residing 
on the boundaries. 


\end{abstract}
\newpage

{\bf 1. Introduction}\\
In Ref. 1, starting from lagrangian field theory and the variational principle, 
it was shown that duality in equations of motion can also be obtained by
introducing explicit spacetime dependence of the lagrangian. Poincare invariance 
was achieved precisely when the duality conditions were satisfied in a particular
way and the same analysis and criteria were valid for both abelian and nonabelian 
dualities.This new  approach to electromagnetic duality also showed 
that some analogue of the holographic principle seems to exist even at length 
scales far larger than that of quantum gravity. The formalism developed in Ref.1
is that of spacetime dependent lagrangians coupled with Schwarz's view$^{5}$  
that in situations with fields not defined everywhere there exist exotic
solutions like monopoles and these solutions are related to duality.

The motivation of the present work (as also that in Ref.1) comes from the
recent discoveries of the behaviour of field theories at the boundaries of 
spacetimes$^{7}$.Specifically, gauge theories have dual description 
in gravity theories in one higher dimension. The theory in higher dimensions
is encoded on the boundary through a different field theory in a lower 
dimension. This discovery of Maldacena and others $^{7}$ is a concrete 
realisation of 't Hooft's {\it holographic principle}$^{7}$.
In this work , the formalism of Ref. 1 is used to study
{\it weak-strong duality}. The first example of weak-strong duality is 
in the seminal works of Coleman and Mandelstam$^{2}$. Coleman showed
that the Sine Gordon (SG) and massive Thirring (MT) models are equivalent order by
order in perturbation theory, provided the coupling constants are related
in a particular way. Mandelstam used an operator approach and showed that
there exists a bosonisation prescription between the theories which lead
to equivalence at the level of the equations of motion, the couplings being
again related in the same way. We mention in passing that duality has been 
studied extensively in the framework of lagrangian field theory $^{3-4}$.  

Let the lagrangian $L'$  be a function of fields $\eta_{\rho}$, 
their derivatives $\eta_{\rho,\nu}$ {\it and the spacetime 
coordinates $x_{\nu}$}, i.e. $L'= L'(\eta_{\rho},\eta_{\rho,\nu}, x_{\nu})$.
Variational principle yields $^{6}$ :
$$\int dV \left(\partial_{\eta}L'
- \partial_{\mu}\partial_{\partial_{\mu}\eta}L'\right) = 0$$
Assuming a separation of variables : 
$L'(\eta_{\sigma},\eta_{\sigma,\nu},.. x_{\nu})
=\Lambda(x_{\nu}) L(\eta_{\sigma},\eta_{\sigma,\nu})$\\
($\Lambda(x_{\nu})$ is the $x_{\nu}$ dependent part and is a finite non-vanishing
function) gives
$$\int dV \left(\partial_{\eta}(\Lambda L)
- \partial_{\mu}\partial_{\partial_{\mu}\eta}(\Lambda L)\right) = 0\eqno(1)$$
Here we consider quantum theories ({\it viz.} SG and MT models) where {\it fields do not 
couple to gravity}. The spacetime dependence will be expressed through $\rho$ 
for the SG model and by $\Lambda$ for the MT model. Under these circumstances,
$\Lambda$ $(\rho)$ is {\it not}  dynamical and is a finite, non-vanishing (operator
and in general complex) function given once and for all at all $x_{\nu}$ 
multiplying the primitive lagrangian $L$. {\it It is like an external field}
and equations of motion for $\Lambda$ $(\rho)$ meaningless. Poincare invariance
and duality invariance is achieved through same behaviour of $\Lambda$ $(\rho)$. 
Specifically, we show that $\Lambda$ $(\rho)$ at infinity is an unitary 
operator and therefore bounded and finite. Within the boundary $\Lambda$ $(\rho)$
is proportional to the identity operator and so ignorable.
The finite behaviour of $\Lambda$ $(\rho)$ on the boundary {\it encodes weak-strong
duality of the SG and MT theories within the boundary}. In this way we are 
reminded of the holographic principle. 
(Finiteness of an operator $A$ in a Hilbert space $H$ is 
understood in the usual sense of its norm. $A$ is bounded if for all vectors
$\vert f> \in H$ one has 
$\vert\vert Af \vert\vert \leq c \vert\vert f\vert\vert$ where $c$ is some number
and $\vert\vert f\vert\vert =<f\vert f>^{1/2}$ ; 
$\vert\vert Af\vert\vert = <f\vert A^{\dagger}A \vert f>^{1/2}$. The norm of 
$A$ is then $\vert\vert A \vert\vert =$ 
lowest upper bound of $\vert\vert Af \vert\vert / \vert\vert f \vert\vert$, 
$\vert f>\not=0$)

{\bf 2.The SG and MT models in $(1+1)$ dimensions}\\
Skyrme $^{2}$ first suggested that the quantum SG solitons although arising from a 
bosonic field theory may be equivalent to fermions interacting through a 
four fermion interaction. Subsequently Coleman$^{2}$ established the 
equivalence within the framework of perturbation theory. The SG and MT models,
both in  $(1 + 1)$ dimensions, are described by the lagrangians:
$$L_{SG} = (1/2) (\partial_{\mu} \phi)(\partial^{\mu}\phi) 
+ m_{0}^ {2}  (m^{2}/\lambda)[cos(\lambda ^{1/2} /m) \phi - 1]$$ 
$$=(1/2) (\partial_{\mu} \phi) (\partial^{\mu} \phi)   
+ (\alpha /\beta^{2}) [cos(\beta \phi) - 1]\eqno(2)$$
$$L_{MT} = i \bar\psi \gamma^{\mu}\partial_{\mu} \psi - m_{F} \bar\psi \psi 
- (1/2) g (\bar\psi \gamma^{\mu} \psi) (\bar\psi \gamma_{\mu} \psi)\eqno(3)$$
where $\phi$ and $\psi$ are bosonic and fermionic fields respectively, 
$\gamma_{\mu}$ are Dirac matrices in $(1+1)$ dimensions,$\alpha  =  m_{0}^{2},
\beta  =  {\lambda}^{1/2}/m$, and normal ordering counterterms have been
absorbed in the parameters $m_{0}^{2}$ and $m_{F}$.
$g_{\mu\nu} \equiv \eta_{\mu\nu} = \pmatrix{1&0\cr 0&-1} ; 
\gamma^{0} = \sigma^{1} = \pmatrix{0&1\cr 1&0}$; 
$\gamma^{1} = i \sigma^{2} = \pmatrix{0&1\cr -1&0} ; 
\gamma^{5} = \gamma^{0}\gamma^{1} = -\sigma^{3}$. 

Our strategy is basically studying three
equations corresponding to the given lagrangian. These are the equations of
motion, the divergence of the current ($j^{\mu}$) and the quantity 
$\partial_{\mu}j^{\nu}+\partial_{\nu}j^{\mu}$  ($\mu \not= \nu$). We will 
show that finiteness of $\Lambda (\rho)$  at infinity imply the original theory 
within the boundary at the level of these equations and is consistent with the
implications of duality both within and on the boundary.
In this sense behaviour of $\Lambda (\rho)$ at the boundary of the theory encode the
original theory within the boundary. Our choice of these particular equations
are dictated by the fact that the first two equations embody all the physics.
This is because we shall start with already renormalised theories.In the case
of the SG theory all divergences that occur in any order of perturbation theory
has been removed by normal ordering and is equivalent to a multiplicative
renormalisation $\alpha$ and an additive renormalisation $\alpha/\beta^{2}$,
$\beta$ is not renormalised. For the MT model, renormalisation implies demanding
that the currents obey proper Ward identities$^{2}$.The third equation acts as a 
consistency check.

The equation of motion of the SG model obtained from $(2)$ is
$$\partial_{\mu}\partial^{\mu}\phi + (\alpha/\beta)sin(\beta\phi)=0\eqno(4a)$$
This is invariant under the transformation $\phi \rightarrow \phi + 2\pi n \beta^{-1}$.
So a topological charge may be defined as (with $\mu ,\nu= 0,1$)
$$Q_{SG} = \int^{\infty}_{-\infty} dx j^{0}=
(-\beta/2\pi)[\phi (x=+\infty,t) - \phi (x=-\infty,t)] = n_{1} - n_{2} = n\eqno(4b)$$
where $n$ is any integer (positive,negative or zero).
$Q_{SG}= +1 (-1)$ for soliton (antisoliton) and soliton-antisoliton bound states
have $Q_{SG}=0$.The associated conserved current is ($\epsilon^{01}=-1$ and $\epsilon^{10}=+1$)
$$j^{\mu} =(-\beta/2\pi) \epsilon^{\mu\nu} \partial_{\nu} \phi \enskip\enskip ;
\enskip\enskip \partial_{\mu} j^{\mu}=0\eqno(4c)$$
and
$$\partial_{0} j^{1} + \partial_{1} j^{0} = (\alpha/2\pi) sin(\beta\phi)\eqno(4d)$$
The equations of motion for the MT model written in terms of the two component
fermion fields are
$$i(\partial_{0}\psi_{1}^{\dagger}-\partial_{1}\psi_{1}^{\dagger})\psi_{1}
-m_{F}\psi_{2}^{\dagger}\psi_{1} - 2 g \psi_{1}^{\dagger}\psi_{1}\psi_{2}^{\dagger}\psi_{2}=0\eqno(5a)$$ 
$$i\psi_{1}^{\dagger}(\partial_{0}\psi_{1}-\partial_{1}\psi_{1})
+m_{F}\psi_{1}^{\dagger}\psi_{2} + 2 g \psi_{1}^{\dagger}\psi_{1}\psi_{2}^{\dagger}\psi_{2}=0\eqno(5b)$$ 
$$i(\partial_{0}\psi_{2}^{\dagger}+\partial_{1}\psi_{2}^{\dagger})\psi_{2}
-m_{F}\psi_{1}^{\dagger}\psi_{2} - 2 g \psi_{1}^{\dagger}\psi_{1}\psi_{2}^{\dagger}\psi_{2}=0\eqno(5c)$$ 
$$i\psi_{2}^{\dagger}(\partial_{0}\psi_{2}+\partial_{1}\psi_{2})
+m_{F}\psi_{2}^{\dagger}\psi_{1} + 2 g \psi_{1}^{\dagger}\psi_{1}\psi_{2}^{\dagger}\psi_{2}=0\eqno(5d)$$ 
The conserved fermionic current is (for nonlocal currents refer to ref.12)
$$k^{\mu}=\bar\psi\gamma^{\mu}\psi \enskip ; \enskip 
\partial_{\mu}k^{\mu} = 0 \enskip ; \enskip                                 
k^{0}=\psi_{1}^{\dagger}\psi_{1}+\psi_{2}^{\dagger}\psi_{2} \enskip ; \enskip
k^{1}=\psi_{2}^{\dagger}\psi_{2}-\psi_{1}^{\dagger}\psi_{1} \eqno(5e)$$
The fermionic charge is 
$Q_{MT} = \int^{\infty}_{-\infty} dx k^{0}=\int^{\infty}_{-\infty} dx \bar\psi \gamma_{0} \psi$.
A single fermion (antifermion) has $Q=1 (-1)$ and bound states of 
the two have $Q=0$, and 
$$\partial_{0}k^{1}+\partial_{1}k^{0}=2im(\psi_{2}^{\dagger}\psi_{1}-\psi_{1}^{\dagger}\psi_{2})\eqno(5f)$$
Coleman's SG theory-MT model (charge-zero sector) equivalence implies
$$(\alpha /\beta^{2})[cos \beta \phi] = -  m_{F} \bar\psi \psi\eqno(6a)$$
$$\beta^{2}/(4\pi) = \pi/(\pi + g)\eqno(6b)$$
$$-(\beta/2\pi)\epsilon^{\mu\nu}\partial_{\nu}\phi = \bar\psi\gamma^{\mu}\psi \equiv j^{\mu}\eqno(6c)$$
Mandelstam's fermionic operator (at any time $t$) construction is
$$\psi_{1}(x)=[c\mu/(2\pi)]^{1/2}e^{\mu/(8\epsilon)}
:e^{-2i\pi\beta^{-1} \int^{x}_{-\infty} dx'(\partial\phi(x')/\partial t)- (i\beta/2) \phi(x) }:\eqno(7a)$$
$$\psi_{2}(x)= - i[c\mu/(2\pi)]^{1/2}e^{\mu/(8\epsilon)}
:e^{-2i\pi\beta^{-1} \int^{x}_{-\infty} dx'(\partial\phi(x')/\partial t)+ (i\beta/2) \phi(x) }:\eqno(7b)$$
where $c\mu$ is an unit of mass (Mandelstam$^{2}$).
The $\psi_{1,2}(x)$ satisfy the Thirring equations of 
motion provided the $\phi$ satisfies the SG equation of motion and 
{\it vice versa}.  
Canonical equal time commutators for $\phi$'s are:
$[\phi(x),\phi(y)]= [\dot{\phi}(x),\dot{\phi}(y)]=0$,
$[\phi(x),\dot{\phi}(y)]=i \delta (x-y)$.
For the $\psi$'s, 
$\{ \psi_{a}(x),\psi_{b}(y) \}=0$ ;\enskip  \enskip 
$\{ \psi_{a}(x),\psi_{b}^{\dagger}(y) \}= z\delta (x-y) \delta_{ab}$
($z$ is another renormalisation constant.) One also has 
$[\phi (y), \psi (x)]= (2\pi/\beta)\theta(x-y)\psi(x)$ for  $x\not=y$.
So, $\psi(x)$ applied to a soliton state with 
$\phi(\infty)-\phi(-\infty)=2\pi/\beta$ reduces it to a state in the vacuum
sector with $\phi(\infty)-\phi(-\infty)=0$. 

{\bf 3.Spacetime dependent lagrangians for SG and MT models}\\
We first consider the SG model and write the modified lagrangian as
$$L'_{SG}=\rho(x,t)L_{SG}\eqno(8)$$
At this point, $L'_{SG}$ need not be hermitian as $\rho^{\dagger}$ need not be
same as $\rho$. Only when $\rho$ is proportional to the identity operator
is $L'_{SG}$ hermitian.
Using $(1)$ and $(2)$ the equations of motion are
$$\dot{\rho} \dot{\phi}-\rho'\phi'
+\rho[\ddot{\phi} - \phi''+ (\alpha/\beta) sin(\beta\phi)] = 0\eqno(9a)$$
$$\dot{\phi}\dot{\rho}^{\dagger}-\phi'\rho'^{\dagger}
+[\ddot{\phi} - \phi''+ (\alpha/\beta) sin(\beta\phi)]\rho^{\dagger}=0\eqno(9b)$$
As $\rho$ is non-dynamical,$(9b)$ will lead to same conclusions as for $(9a)$.
So we confine ourselves to $(9a)$. 
The symmetries of $(9)$ are still $\phi\rightarrow - \phi$ and
$\phi\rightarrow \phi + 2\pi n \beta^{-1}$. In presence of $\rho$ define the
new topological current as 
$$j^{\mu}_{\rho}= (-\beta/2\pi) [\epsilon^{\mu\nu}\rho(\partial_{\nu}\phi)
+\epsilon^{\mu\nu}(\partial_{\nu}\rho) \phi]\eqno(10a)$$
Then $\partial_{\mu}j^{\mu}_{\rho}=0$ and the new conserved 
topological charge is given by 
$$Q_{SG\rho}= \int^{\infty}_{-\infty} dx j^{0}_{\rho}
=(-\beta/2\pi)[\rho(x=+\infty,t)\phi (x=+\infty,t)$$ 
$$-\rho(x=-\infty,t) \phi (x=-\infty,t)]
= n_{1\rho} - n_{2\rho} = n_{\rho}\eqno(10b)$$
where $n_{\rho}$ is again some integer.
In presence of $\rho$ ,$(4d)$ takes the form:
$$\partial_{0} j^{1}_{\rho} + \partial_{1} j^{0}_{\rho}
=(\beta/2\pi)[\rho\partial_{\mu}\partial^{\mu}\phi-
(\partial_{\mu}\partial^{\mu}\rho)\phi]
+ (\alpha\rho/\pi)sin(\beta\phi)$$
$$=(\alpha\rho/2\pi)sin(\beta\phi)
-(\beta/2\pi)(\partial_{\mu}\partial^{\mu}\rho)\phi\eqno(11)$$
where we have imposed the fact that at $\infty$ 
$(4a)$ must be valid.
$(11)$ reduces to $(4d)$ for $\rho$ proportional to the {\it identity operator}
(within the boundary).

We want to establish that the finiteness of $\rho$
on the boundary implies the usual duality between SG and MT models within the 
boundary. This means that there is a  solution for 
$\rho$ at $x \rightarrow \infty$, which in the weak coupling limit 
(i.e.$\beta\rightarrow 0 $ which always implies $\beta^{2}\rightarrow 0$; we 
shall always invoke this stronger condition) implies that $(9)$
reduces to $(4a)$ and $(11)$ reduces to $(4d)$. Finiteness of $\rho$ will be 
shown to be equivalent to the fact that $\rho$ in the above limit is
proportional to an unitary operator so that
$\rho^{\dagger}\rho(=\rho\rho^{\dagger})$ is proportional to the identity
operator.We now show that there exists a solution for $\rho$ whose behaviour
at $x\rightarrow \infty$ satisfies all these conditions.Recall $(7a),(7b)$.
We take $\rho$ to be precisely these wthout the normal ordering.
$$\rho_{1}(x)=[c\mu/(2\pi)]^{1/2}e^{\mu/(8\epsilon)}
e^{-2i\pi\beta^{-1} \int^{x}_{-\infty} dx' \dot{\phi}(x')- (i\beta/2) \phi(x) }$$
$$\rho_{2}(x)= - i[c\mu/(2\pi)]^{1/2}e^{\mu/(8\epsilon)}
e^{-2i\pi\beta^{-1} \int^{x}_{-\infty} dx'\dot{\phi}(x')+ (i\beta/2) \phi(x) }$$
We now estimate $\rho(\infty,t)$, i.e. for $x\rightarrow\infty$. The integral in
the exponent of $\rho$ now becomes 
$\int^{\infty}_{-\infty} dx \dot{\phi}(x)$.
The integrand is an operator. So the integral has to be understood in terms of
its {\it expectation value} $^{8}$. Then this integral from $-\infty\enskip to\enskip +\infty$
will be dominated by the classical value of $\phi$  i.e. $^{8}$
$$\phi_{cl} = (4/\beta) tan^{-1}[ e ^ {m(x-x_{0}-ut)(1-u^{2})^{-1/2}}]\eqno(12)$$
$u$ is velocity boost on static soliton solution.
Integrating for some fixed time $t$ gives 
$$\rho_{1}(\infty,t)\Rightarrow A e^{(4\pi ^{2}/\beta^{2})iu 
- (i\beta/2)\phi(\infty,t)}\eqno(13a)$$
$$\rho_{2}(\infty,t)\Rightarrow-i A e^{(4\pi ^{2}/\beta^{2})iu 
+ (i\beta/2)\phi(\infty,t)}\eqno(13b)$$
where $A=[c\mu/(2\pi)]^{1/2}e^{\mu/(8\epsilon)}$. So the first term in the 
exponent of equation $(13)$ ({\it viz.} $(4\pi ^{2}/\beta^{2})iu$)
is proportional to the identity operator and 
$\rho^{\dagger}\rho(=\rho\rho^{\dagger})$ is proportional to the 
identity operator (modulo $A^{2}$).

First consider the equation of motion {\it viz.} $(9a)$ for (say $\rho_{1}$). 
$\dot{\rho_{1}}=-(i\beta/2)\{\dot{\phi}, \rho_{1}\}$ where $\{ ,\}$ denotes the 
anticommutator. The commutator  
$[\dot{\phi}(x,t),\rho_{1}(y,t)]=-(\beta/2)\rho_{1}(y,t)\delta(x-y) \enskip ; y\rightarrow\infty$.
Then, $\dot{\rho_{1}} = -i(\beta^{2}/4 + \beta\dot{\phi})\rho_{1}$ so that 
$\dot{\rho_{1}} \dot{\phi}$ is of  $O(\beta)$ and higher. Similarly, $\rho_{1}'\phi'$ is also
of $O(\beta)$ and higher. Hence the first two terms are negligible compared to the
third term which is of $O(\beta^{0})$ for small $\beta$ 
($sin (\beta\phi)\simeq \beta\phi$ for small $\beta$). Therefore, the behaviour of
$\rho_{1}(\infty,t)$ implies the usual equation of motion within the boundary
for small $\beta$.Same is true for $\rho_{2}$.

Next the term $(\beta/2\pi)(\rho_{1}''- \ddot{\rho_{1}})\phi$ in $(11)$
is at least $O(\beta^{2})$ as 
$\ddot{\rho_{1}}=-[\beta^{4}/16+(\beta^{3}/2)\dot{\phi}+   
\beta^{2}\dot{\phi}^{2}+i\beta\ddot{\phi}]\rho_{1}$,
and
$\rho_{1}''=[-(i\beta/2)\phi''-(\beta^{2}/4)(\phi')^{2}]\rho_{1}$.
So this is ignorable. Again we impose that duality holds on the 
boundary. So $(4a)$ and $(4d)$ holds and therefore 
$\rho_{1}(\alpha/2\pi)sin(\beta\phi)
=\rho_{1} (\partial_{0} j^{1} + \partial_{1} j^{0})$.Thus on the boundary
$\partial_{0} j^{1}_{\rho_{1}} + \partial_{1} j^{0}_{\rho_{1}}
=\rho_{1}(\partial_{0} j^{1} + \partial_{1} j^{0})$,
and the norm of $\rho$ is unity whereas within the boundary $\rho$
is proportional to the identity operator.
Same conclusions hold for $(9b)$.
Therefore $\rho^{\dagger}\rho(=\rho\rho^{\dagger})$ being
proportional to the identity operator at $x\rightarrow\infty$ implies duality
between MT and SG models {\it within the boundary}. 

Now consider the MT model with spacetime dependence. 
$$L'_{MT}=\Lambda(x,t)L_{MT}\eqno(14)$$
The equations of motion (using $(1)$ and $(3)$) are
$$i(\Lambda^{\dagger}\partial_{0}\Lambda-\Lambda^{\dagger}\partial_{1}\Lambda)
\psi_{1}^{\dagger}\psi_{1}
+(\Lambda^{\dagger}\Lambda)[i(\partial_{0}\psi_{1}^{\dagger}-\partial_{1}\psi_{1}^{\dagger})\psi_{1}
-m_{F}\psi_{2}^{\dagger}\psi_{1} - 2 g \psi_{1}^{\dagger}\psi_{1}\psi_{2}^{\dagger}\psi_{2}]
=0\eqno(15a)$$ 
$$i\psi_{1}^{\dagger}\psi_{1} (\Lambda\partial_{0}\Lambda^{\dagger}-\Lambda\partial_{1}\Lambda^{\dagger})
+[i\psi_{1}^{\dagger}(\partial_{0}\psi_{1}-\partial_{1}\psi_{1})
+m_{F}\psi_{1}^{\dagger}\psi_{2} 
+ 2 g \psi_{1}^{\dagger}\psi_{1}\psi_{2}^{\dagger}\psi_{2}](\Lambda^{\dagger}\Lambda)
=0\eqno(15b)$$ 
$$i(\Lambda\partial_{0}\Lambda^{\dagger}+\Lambda\partial_{1}\Lambda^{\dagger})
\psi_{2}^{\dagger}\psi_{2}
+(\Lambda^{\dagger}\Lambda)[i\psi_{2}^{\dagger}(\partial_{0}\psi_{2}+\partial_{1}\psi_{2})
+m_{F}\psi_{2}^{\dagger}\psi_{1} + 2 g \psi_{1}^{\dagger}\psi_{1}\psi_{2}^{\dagger}\psi_{2}]
=0\eqno(15c)$$ 
$$i\psi_{2}^{\dagger}\psi_{2}(\Lambda^{\dagger}\partial_{0}\Lambda+\Lambda^{\dagger}\partial_{1}\Lambda)
+[i(\partial_{0}\psi_{2}^{\dagger}+\partial_{1}\psi_{2}^{\dagger})\psi_{2}
-m_{F}\psi_{1}^{\dagger}\psi_{2} - 2 g \psi_{1}^{\dagger}\psi_{1}\psi_{2}^{\dagger}\psi_{2}](\Lambda^{\dagger}\Lambda)
=0\eqno(15d)$$ 
If we {\it identify $\Lambda$ with $\rho$, the first
terms of each of the equations $(15)$ are of order at least $\beta$ for 
$x\rightarrow\infty$; wheras the second terms are of  $O(\beta^{0})$
(since $\Lambda^{\dagger}\Lambda=1$). Hence the usual MT model equations of
motion $(5)$ are recovered for $x\rightarrow\infty$.}
With this hindsight, $\Lambda^{\dagger}\Lambda$ will be a bilinear in $\psi$'s 
(we have shown that $\Lambda$ can be identified with $\rho$). So the above four
equations can be re-written by commuting the $\Lambda^{\dagger}\Lambda$ 
and $\psi^{\dagger}\psi$ terms through the other terms:
$$i(\Lambda^{\dagger}\partial_{0}\Lambda-\Lambda^{\dagger}\partial_{1}\Lambda)
\psi_{1}^{\dagger}\psi_{1}
+(\Lambda^{\dagger}\Lambda)[i(\partial_{0}\psi_{1}^{\dagger}-\partial_{1}\psi_{1}^{\dagger})\psi_{1}
-m_{F}\psi_{2}^{\dagger}\psi_{1} - 2 g \psi_{1}^{\dagger}\psi_{1}\psi_{2}^{\dagger}\psi_{2}]
=0\eqno(16a)$$ 
$$i(\Lambda\partial_{0}\Lambda^{\dagger}-\Lambda\partial_{1}\Lambda^{\dagger})\psi_{1}^{\dagger}\psi_{1}
+(\Lambda^{\dagger}\Lambda)[i\psi_{1}^{\dagger}(\partial_{0}\psi_{1}-\partial_{1}\psi_{1})
+m_{F}\psi_{1}^{\dagger}\psi_{2} 
+ 2 g \psi_{1}^{\dagger}\psi_{1}\psi_{2}^{\dagger}\psi_{2}]
=0\eqno(16b)$$ 
$$i(\Lambda\partial_{0}\Lambda^{\dagger}+\Lambda\partial_{1}\Lambda^{\dagger})
\psi_{2}^{\dagger}\psi_{2}
+(\Lambda^{\dagger}\Lambda)[i\psi_{2}^{\dagger}(\partial_{0}\psi_{2}+\partial_{1}\psi_{2})
+m_{F}\psi_{2}^{\dagger}\psi_{1} + 2 g \psi_{1}^{\dagger}\psi_{1}\psi_{2}^{\dagger}\psi_{2}]
=0\eqno(16c)$$ 
$$i(\Lambda^{\dagger}\partial_{0}\Lambda+\Lambda^{\dagger}\partial_{1}\Lambda)\psi_{2}^{\dagger}\psi_{2}
+(\Lambda^{\dagger}\Lambda) [i(\partial_{0}\psi_{2}^{\dagger}+\partial_{1}\psi_{2}^{\dagger})\psi_{2}
-m_{F}\psi_{1}^{\dagger}\psi_{2} - 2 g \psi_{1}^{\dagger}\psi_{1}\psi_{2}^{\dagger}\psi_{2}]
=0\eqno(16d)$$ 
The conserved currents (using $(16)$) are then given by
$$k^{\mu}_{\Lambda}=\Lambda^{\dagger}\Lambda\bar\psi\gamma^{\mu}\psi
\enskip ; \enskip  \partial_{\mu}k^{\mu}_{\Lambda}=0
\enskip ; \enskip k^{0}_{\Lambda}=\Lambda^{\dagger}\Lambda k_{0}
\enskip ; \enskip k^{1}_{\Lambda}=\Lambda^{\dagger}\Lambda k_{1}\eqno(17a)$$  
$$\partial_{0}k^{1}_{\Lambda}+\partial_{1}k^{0}_{\Lambda}
=(\Lambda^{\dagger}\Lambda)2im(\psi_{2}^{\dagger}\psi_{1}-\psi_{1}^{\dagger}\psi_{2})$$
$$+\partial_{0}(\Lambda^{\dagger}\Lambda)(\psi_{2}^{\dagger}\psi_{2}-\psi_{1}^{\dagger}\psi_{1})
+\partial_{1}(\Lambda^{\dagger}\Lambda)(\psi_{1}^{\dagger}\psi_{1}+\psi_{2}^{\dagger}\psi_{2})\eqno(17b)$$
$(17b)$ is the analogue of $(5f)$. It is trivial to se that $(5f)$ is recovered
from $(17b)$ for $x\rightarrow\infty$ (since $\Lambda^{\dagger}\Lambda=1$ at $\infty$).
{\it Therefore, weak-strong duality of the SG and 
MT models can be re-expressed by  stating that the spacetime dependence for both
theories are given by the same function. The finite behaviour of this function
at $x\rightarrow\infty$ encodes the duality of the theories within the boundary.}

{\bf 4. Possibility of Noncommuting Coordinates on the boundary}\\ 
We now explore the possibility of constructing noncommuting coordinates on the
boundary.(Recent references are in ref.9).
With respect to bare particle creation and annihilation operators the
SG field can be written $^{2}$ as
$$\phi(x,t)=\phi^{+}(x,t) + \phi^{-}(x,t)\eqno(18)$$
where $\phi^{+}$ and $\phi^{-}$ satisfy the commutation relations
$$[\phi^{+}(x,t+\delta t),\phi^{-}(y,t)]=\bigtriangleup_{+}((x-y)^{2} -(\delta t+i\epsilon)^{2})\eqno(19a)$$
For small separations $(\delta x)$ 
$$\bigtriangleup_{+}((\delta x)^{2} -(\delta t+i\epsilon)^{2})
= - (4\pi)^{-1}ln[c^{2}\mu ^{2}((\delta x)^{2}-(\delta t+i\epsilon)^{2})]+O((\delta x)^{2})\eqno(19b)$$
Consider two infinitesimally close points $x_{1}$ and $x_{2}$ near the spatial
boundary. So $x_{1}, x_{2}\rightarrow\infty$ and $x_{1}- x_{2}=\delta x$.We shall
confine our discussions to $\rho_{1}$. At infinity 
$\rho_{1}(\infty,t)\Rightarrow A e^{(4\pi ^{2}/\beta^{2})iu 
- (i\beta/2)\phi(\infty,t)}$
Let 
$$\rho_{1}(x_{1},t)= A e^{(4\pi ^{2}/\beta^{2})iu 
- (i\beta/2)\phi(x_{1},t)}=Ae^{iz^{1}}\equiv \Omega^{1}\eqno(20a)$$
$$\rho_{1}(x_{2},t+\delta t)= A e^{(4\pi ^{2}/\beta^{2})iu 
- (i\beta/2)\phi(x_{2},t+\delta t)}=Ae^{iz^{2}}\equiv \Omega^{2}\eqno(20b)$$
where $z_{1},z_{2}$ are like two different points on some circle with the
(one dimensional) angular coordinate $z$. Then
$$\Omega^{1}\Omega^{2}=e^{-[z^{1},z^{2}]} \Omega^{2}\Omega^{1}\eqno(21)$$
and 
$$[z^{1},z^{2}]=(\beta^{2}/8\pi)ln[1+2i\epsilon
\delta t ((\delta x)^{2}-(\delta t)^{2})^{-1}]
=i\Theta^{12}\eqno(22)$$
with 
$$\Theta^{12}={\epsilon(\delta t)(\beta^{2}/4\pi)\over(\delta x)^{2}-(\delta t)^{2}}\eqno(23)$$
Here $\delta t=t_{1}-t_{2}$. We have neglected $\epsilon^{2}$ and 
$\epsilon^{2}(\delta t)^{2}$ terms and have taken the principal value of the logarithm.
Denominator in $(23)$ is essentially the metric. If we assume it is 
always non-null and further that $t_{1}\not=t_{2}$, then $\Theta^{12}$ is 
antisymmetric with respect to interchange of the indices $1,2$. So
here we have a structure reminiscent of noncommuting coordinates along
the lines described in ref.10. If $z$ is given the staus of a
one-dimensional coordinate then {\it different points on it are non-commuting}.
This is as far as we can go because the SG and MT theories are
$1(space)+1(time)$ dimensional theories. Moreover, one should study the implications
of ref.11 that perturbative unitarity of the $S$-matrix 
is possible for both space noncommutativity as well as light-like noncommutativity.
Keeping all these facts
in mind we can still say that our formalism can accommodate structures that
reminds one of noncommuting coordinates. Note that the factor $(\beta^{2}/4\pi)$
in $(23)$ can be replaced by $\pi/(\pi + g)$ if we invoke the duality of the
SG and MT models. {\it Therefore, the weak-strong
duality of the SG and MT models can be restated in our formalism by saying 
that they give rise to the same noncommuting structure of coordinate like objects
on the boundary. If we assume that the Mandelstam vertex operator construction
is unique then our spacetime dependent functions ($\rho,\Lambda$)
and all subsequent results are also unique.}

{\bf 5. Conclusion }\\ 
We have applied our formalism of spacetime dependent
lagrangians developed in Ref.1 to the weak-strong duality of the SG
and MT models. We have shown that our formalism consistently
describes both  (classical) electromagnetic duality and (quantum) weak-strong
duality. We have also shown that on the boundary one can construct objects 
that encode the duality of the theories {\it within the boundary}.
There are  also possibilities for constructing noncommutative coordinates.
Our conjecture in Ref.1 that 't Hooft's holographic principle has
analogues in length scales much larger than quantum gravity has also been
illustrated. The essential principle seems to be the existence of
duality in the theories under consideration.

\end{document}